\documentstyle[preprint,aps]{revtex}
\begin{document}
\draft
\preprint{UTPT-97-07}
\title{DO SUPERMASSIVE BLACK HOLES EXIST AT THE CENTER OF GALAXIES?}
\author{J. W. Moffat}
\address{Department of Physics, University of Toronto,
Toronto, Ontario M5S 1A7, Canada}

\date{\today}
\maketitle

\begin{abstract}%
Models of superdense star clusters at the center of galaxies are investigated to see whether
such objects can be stable and long-lived based on evaporation and collision time-scales and
stability criteria. We find that physically reasonable models of massive clusters of stellar
remnants can exist with masses $\geq 10^6\,M_{\odot}$, which could simulate black holes at 
the center of galaxies with large $M/L$ ratios and gas motions of order $\geq 10^3$ km
$s^{-1}$. It follows that the evidence is not conclusive for massive dark objects at the center of
galaxies being black holes.
\end{abstract}
\vskip 0.2 true in
{\it Subject headings}: black hole physics - galaxies: individual (NGC 4258) - galaxies: nuclei -
masers

\pacs{ }

\section{Introduction}

There has been a recent show of exuberance about the supporting evidence for the
existence of $\sim 10^6-10^{9.5}\,M_{\odot}$ supermassive black holes at the center of
galaxies 
(Kormendy 1993; Kormendy \& Richstone 1995 
and references therein, Kormendy et al., 1997). However, as emphasized by Kormendy and
Richstone, large
$M/L$ ratios and gas motions of order $\geq 10^3$ km $s^{-1}$ do not provide a unique
signature for supermassive black holes. Could these central dark objects in galaxies be
massive clusters of stellar remnants, brown dwarfs, or halo dark matter? The arguments used to
support the evidence that the massive dark objects are supermassive black
holes are of an indirect astrophysical nature. The event horizons of the black holes would reside
$\sim 10^4-10^5$ Schwarzschild radii below the spatial resolution of the Hubble space
telescope. 

Maoz (1995) has
used the measurement of the rotation curve of maser emission sources at the center of NGC 4258
(Miyoshi et al., 1995) to support the
conclusion that this dynamical system cannot be a central cluster, unless the cluster consists of 
extremely dense objects with mass $\leq 0.03\,M_{\odot}$, e.g., low-mass black holes or
elementary particles.

We shall show that it is possible to construct models of clusters with central masses in excess of 
$10^6\,M_{\odot}$, which have reasonably long evaporation and collision time-scales for
such galaxies and can be expected to be stable objects. Miyoshi et al., obtained data using the
Very Long Baseline Array, confirming that for NGC 4258 ($M\approx
4\times10^7\,M_{\odot})$ the rotation curve  is Keplerian to a high precision. If the
rotation is circular, then the mass interior to $0''.005=0.18$ pc is $M\sim 4\times
10^7\,M_{\odot}.$ This gets closer than is normally possible to a potential black hole at the
center.

Supermassive stars with arbitrarily large redshifts were first studied by Zel'dovich and
Poduretz (1965) and subsequently by Ipser \& Thorne (1968), Bisnovatyi-Kogan \& Zel'dovich
(1969), Ipser (1969, 1970), Fackerell (1970), Bisnovatyi-Kogan \& Thorne
(1970) and by Weinberg (1972). These studies were motivated by the suggestion of Hoyle and
Fowler (1967) and Zapolsky (1968) that the emission lines of QSOs might come from the centers
of supermassive star
clusters, whose gravitational fields would produce all or most of the observed redshift. Since then
evidence strongly supports that QSOs are at cosmological distances and that the redshifts are due
to the expansion of the universe. However, there is now a renewed interest to study dark massive
objects at the center of galaxies, because of the possibility that they could be either black holes or 
supermassive clusters.

\section{Equilibrium Configurations of Supermassive Clusters}

Let us give an overview of the supermassive cluster equilibrium problem.  We shall assume that 
the core cluster is mainly supported by the pressure of radiation rather than of matter and that it is
in convective equilibrium and has a uniform chemical composition. For radiation the energy 
density is $\epsilon\sim 3p$ and the cluster can be described by a Newtonian polytrope with 
\begin{equation}
\epsilon=\rho-mn=(\gamma-1)^{-1}p,
\end{equation}
where $m$ and $n$ denote the mass of a star in the cluster and the number density of stars,
respectively, $\rho$ and $p$ denote the mass density and pressure, respectively, and
$1/(\gamma-1)$ is a constant of proportionality. The condition of uniform
entropy per star then yields the polytrope equation of state
\begin{equation}
p=K\rho^{\gamma},
\end{equation}
where $K$ is a constant of proportionality. The standard polytrope equilibrium equation is
(Chandrasekhar 1939):
\begin{equation}
\frac{1}{\xi^2}\frac{d}{d\xi}\xi^2\frac{d\theta}{d\xi}+\theta^{1/(\gamma-1)}=0,
\end{equation}
where the radial variable $r$ is related to $\xi$ by
\begin{equation}
r=\biggl(\frac{K\gamma}{4\pi G(\gamma-1)}\biggr)^{1/2}\rho_0^{(\gamma-2)/2}\xi,
\end{equation}
and
\begin{mathletters}
\begin{eqnarray}
\rho&=&\rho_0\theta^{1/(\gamma-1)},\\
p&=&K\rho_0^{\gamma}\theta^{\gamma/(\gamma-1)}.
\end{eqnarray}
\end{mathletters}
Boundary conditions on the Lane-Emden function $\theta$ are given by $\theta_0=1$ and
$\theta'_0=0$.

The total energy for the core Newtonian cluster is $E=T+V$, where $T$ and $V$ denote the
thermal energy and gravitational potential energy, respectively, and 
for Newtonian polytropes the total energy is $E=-(3\gamma-4)GM^2/(5\gamma-6)R$.

Since we have assumed that the supermassive cluster is mainly supported by thermal radiation,
we can choose $\gamma\sim 4/3$ and its mass is given by (Chandrasekhar 1939):
\begin{equation}
\label{totalmass}
M=25.3620\biggl(\frac{K}{\pi G}\biggr)^{3/2}
\end{equation}
and its radius by
\begin{equation}
\label{radius}
R=6.89685\biggl(\frac{K}{\pi G}\biggr)^{1/2}\rho_0^{-1/3},
\end{equation}
where $\rho_0$ is the central density.

The structure of the supermassive cluster can be described by a Newtonian polytrope with 
$\gamma\sim 4/3$ (for NGC4258 $GM/R\sim 1.6\times 10^{-5}$ which is locally Newtonian),
but to settle the question of stability we need General Relativity (GR), for a cluster with 
$\gamma\sim 4/3$ is sensitively balanced between stability and instability, so small effects due
to GR and matter pressure must be accounted for, although they play little role in the structure
calculations (Chandrasekhar 1964).

We characterize the spherically symmetric cluster by a perfect fluid with the
metric (Weinberg 1972):
\begin{equation}
ds^2=B(r)dt^2-A(r)dr^2-r^2(d\theta^2+\sin^2\theta d\phi^2),
\end{equation}
where
\begin{mathletters}
\begin{eqnarray}
A(r)&=&\biggl[1-\frac{2GM(r)}{r}\biggr]^{-1},\\
B(r)&=&\exp\biggl\{-2G\int^{\infty}_r\frac{dr}{r^2}[M(r')+4\pi r^{'3}p(r')]
\biggl[1-\frac{2GM(r')}{r'}\biggr]^{-1}\biggr\}.
\end{eqnarray}
\end{mathletters}
Moreover, we have
\begin{equation}
M(r)=4\pi\int_0^rdr'r^{'2}\rho(r').
\end{equation}
Outside the cluster $p(r)$ and $\rho(r)$ vanish and the metric is descibed the Schwarzschild
solution
\begin{equation}
B(r)=A^{-1}(r)=1-\frac{2G M(R)}{r}\quad {\rm for}\, r\geq R.
\end{equation}

The general relativistic equation is given by the Oppenheimer-Volkoff
equation (Oppenheimer \& Volkoff 1939):
\begin{equation}
p'(r)=-\frac{GM(r)\rho(r)}{r^2}\biggl[1+\frac{p(r)}{\rho(r)}\biggr]
\biggl[1+\frac{4\pi r^3p(r)}{M(r)}\biggr]\biggl[1-\frac{2GM(r)}{r}\biggr]^{-1}.
\end{equation}
The thermal and gravitational energies are given by
\begin{mathletters}
\begin{eqnarray}
T&=&4\pi\int^R_0drr^2\biggl[1-\frac{2GM(r)}{r}\biggr]^{-1/2}\epsilon(r),\\
V&=&4\pi\int_0^Rdrr^2\biggl\{1-\biggl[1-\frac{2GM(r)}{r}\biggr]^{-1/2}\biggr\}
\rho(r),
\end{eqnarray}
\end{mathletters}
where $\epsilon(r)$ denotes the internal energy density.

Expanding in powers of $GM(r)/r$ we get
\begin{mathletters}
\begin{eqnarray}
T&=&4\pi\int_0^Rdrr^2\biggl[1+\frac{GM(r)}{r}+...\biggr]\epsilon(r),\\
V&=&-4\pi\int_0^Rdrr^2\biggl[\frac{GM(r)}{r}+\frac{3G^2M^2(r)}{2r^2}+...\biggr]\rho(r).
\end{eqnarray}
\end{mathletters}

A cluster described by a perfect fluid may pass from stability to instability with a
radial normal mode at a value of the central density $\rho_0$ for which the energy $E$ and the
number of stars $N$ is stationary (Chandrasekhar 1964):
\begin{mathletters}
\begin{eqnarray}
\frac{\partial E}{\partial\rho_0}&=&0,\\
\frac{\partial N}{\partial\rho_0}&=&0.
\end{eqnarray}
\end{mathletters}
The energy $E$ has the approximate form 
\begin{eqnarray}
E\simeq 4\pi\int_0^Rdrr^2\epsilon(r)+4\pi G\int_0^RdrrM(r)\epsilon(r)
-4\pi G\int_0^RdrrM(r)\rho(r)\nonumber\\
-6\pi G^2\int_0^RdrM^2(r)\rho(r).
\end{eqnarray}
The total pressure is $p=p_r+p_m$ where $p_r$ and $p_m$ denote the radiation and matter
pressure, respectively, and the internal energy density is
\begin{equation}
\epsilon=3p_r\biggl[1+\frac{\beta}{3(\Gamma-1)}\biggr],
\end{equation}
where $\Gamma$ is the specific heat ratio of the matter and $\beta=p_m/p_r$. The total 
pressure is $p=p_r(1+\beta)$, so that to first order in small $\beta$, the ratio of energy density 
to pressure gives
\begin{equation} 
\epsilon\simeq 3p\biggl[1-\frac{(3\Gamma-4)}{3(\Gamma-1)}\beta+O(\beta^2)\biggr].
\end{equation}
We now integrate by parts and to first order in $\beta$ and employing the approximate
Newtonian equation
\begin{equation}
p'(r)\simeq -\frac{GM(r)\rho(r)}{r^2}
\end{equation}
to evaluate $\rho,p$ and $M(r)$, we get (Weinberg 1972):
\begin{equation}
E\simeq -\frac{(3\Gamma-4)}{2(\Gamma-1)}\beta\frac{GM^2}{R}+5.1\frac{G^2M^3}{R^2}.
\end{equation}

The stability criterion
\begin{equation}
\frac{\partial E}{\partial r}=\frac{\partial E}{\partial\rho_0}\frac{\partial\rho_0}{\partial r}=0
\end{equation}
yields the minimum radius for stability
\begin{equation}
R_{\rm min}=\frac{20.4(\Gamma-1)}{(3\Gamma-4)}\frac{GM}{\beta}.
\end{equation}
For a core cluster with $M=3.6\times 10^7\,M_{\odot}, \beta\sim 0.1$ and
$\Gamma\sim 5/3$ we get the minimum radius
\begin{equation}
R_{\rm min}=2.3\times 10^{-4}\,{\rm pc},
\end{equation}
which is deep inside a typical supermassive core residing in a galaxy (NGC4258).
The ratio of $R_{\rm min}$ to the Schwarzschild radius $R_S=2GM$ is
$R_{\rm min}/R_S=6.8/\beta$.

Although we have used a perfect fluid to investigate the stability properties of a supermassive
cluster, we consider that it is a good approximation and that it is safe to conclude that a
kinetic particle description of the supermassive cluster will yield similar results.

\section{Timescales of Evaporation and Collisions}

The time-scale of evaporation of a bound system of objects with a single mass can be
determined to be $t_{\rm evap}\approx 300t_{\rm relax}$ (Spitzer \& Thuan 1972; 
Spitzer \& Hart 1971; Binney \& Tremaine 1987) with
\begin{equation}
t_{\rm relax}=\biggl[\frac{0.14N}{{\rm ln}\,(0.4N)}\biggr]\biggl(\frac{R^3_{1/2}}
{GM}\biggr)^{1/2},
\end{equation}
where $N$ is the number of objects in the cluster, $M$ is the cluster mass, and
$R_{1/2}$ is the cluster radius within which lies half of the cluster's mass. 

Let us assume that all the mass in the dense cluster is confined within a spherically
symmetric core, mantle and halo. The core is described by (Bisnovatyi-Kogan \& Thorne
1970):
\begin{equation}
\rho_c(r)=\rho_0\biggl[1-\frac{2\pi G\rho_0r^2}{3\gamma}\biggr],
\end{equation}
where $\rho_0$ is the central density. This core is joined smoothly to a mantle with the density
profile
\begin{equation}
\rho_m(r)=\biggl(\frac{\gamma}{1+6\gamma+\gamma^2}\biggr)
\biggl(\frac{1}{2\pi Gr^2}\biggr)
\end{equation}
with the join point in a region just outside the core radius,
\begin{equation}
\label{join}
r_c=\biggl(\frac{\gamma}{2\pi\rho_0 G}\biggr)^{1/2}.
\end{equation}
Outside the mantle a Newtonian envelope is constructed which is assumed to be convectively
stable and has a finite radius $r_e$ at which $\rho$ goes to zero in a polytropic fashion,
\begin{equation}
\rho_e(r)\propto (r_e-r)^N,\quad N>0.
\end{equation}

We now assume that almost all the mass is confined within the core and the mantle.
Then, we have $\rho\approx \rho_{\rm c} +\rho_{\rm m}$ and the mass is
\begin{equation}
\label{mass}
M\approx 4\pi\biggl[\int^{r_c}_0drr^2\rho_c(r)
+\int^{r_m}_{r_c} drr^2\rho_m(r)\biggr]
\approx\frac{4\pi}{3}r^3_c\rho_0 +\frac{8\pi^2G\rho_0^2}{15\gamma}r_c^5
+\frac{2\gamma}{G}(r_m-r_c).
\end{equation}
From (\ref{join}) and with $\gamma <1$ and $r_m >> r_c$, it follows that we can ignore the
first
two terms on the right-hand side of (\ref{mass}). The evaporation time for $R_{1/2}\sim 0.2$
pc,
$M\sim 3.6\times 10^7\,M_{\odot}$ and a dense cluster
consisting of neutron stars with $m\sim 1.4\,M_{\odot}$ is
\begin{equation}
\label{evaporation}
t_{\rm evap}\sim 10\,{\rm Gyr},
\end{equation}
which is an adequate lifetime for the superdense cluster.

Maoz (1995) has imposed constraints on the mass distribution of a dense stellar cluster by
using the observational findings for NGC 4258. The high-velocity  maser emission data obtained
for NGC 4258 describe a nearly planar structure, and the velocity decreases from $v_{\rm
in}=1080\pm2\,{\rm km}\,s^{-1}$ at a distance $r_{\rm in}=0.13$ pc to $v_{\rm out}=770\pm
2\,{\rm km}\,s^{-1}$ at a distance $r_{\rm out}=0.25$ pc (Miyoshi et al., 1995). Assuming 
circular motion and a perfectly planar disk, it was found that it can be fitted very well by a
Keplerian relation. The systematic deviation of the velocity profile from a Keplerian relation is
$\Delta v \leq 3\,{\rm km}\,s^{-1}$ (Maoz 1995) or a fractional deviation of $\Delta v/{\bar
v}\leq 4\times 10^{-3}$, where ${\bar v}$ is the average rotational velocity. Maoz assumed that
the entire mass is within a radius $r_{\rm in}$ with a mass density profile described by a
Plummer model (Binney \& Tremaine 1987):
\begin{equation}
\label{Plummer}
\rho(r)=\rho_0\biggl[1+\frac{r^2}{r^2_c}\biggr]^{-5/2},\quad
M(<r)=\frac{4\pi\rho_0}{3}r^3\biggl(1+\frac{r^2}{r_c^2}\biggr)^{-3/2}.
\end{equation}
The ratio of the cluster mass enclosed between the spheres of radii $r_{\rm in}$ and $r_{\rm
out}$ to its mass within $r_{\rm in}$ is given by
\begin{equation}
\label{fraction}
\delta_{\rm Kep}=[M(<r_{\rm out})-M(<r_{\rm in})]/M(<r_{\rm in}).
\end{equation}
Then, for $\delta_{\rm Kep}\sim 0.01$ and solving for $r_c$ using (\ref{Plummer}) and
(\ref{fraction}), it follows that $r_c\leq 0.012$ pc and
with $R_{1/2}=1.3r_c$ and $r_{\rm in} \gg r_c$,
we get for a hypothetical cluster of neutron stars ($m\sim 1.4\,M_{\odot}$) at the center of
NGC 4258 the evaporation time $t_{\rm evap}\sim 10^8$ yr. Since this is a period of time
much shorter than the age of the galaxy, Maoz ruled out the possibility that NGC 4258 is a
cluster of stars with mass $\approx 1.4\,M_{\rm \odot}$. A cluster of stars with mass $\sim
0.03\,M_{\odot}$ would yield $t_{\rm evap}\sim 6$ Gyr, which would not be ruled out, but
would be difficult to reconcile with collision timescales, unless the objects are extremely dense,
e.g., light black holes or elementary particles, which are difficult to reconcile with any known
theory of structure formation or stellar evolution.

Consider now our model of a superdense cluster of stars with a mass profile given by
(\ref{mass}). Assuming that $r_c\ll r_{\rm in}\sim r_m$ we get 
\begin{equation}
r_{\rm in}\leq \frac{r_{\rm out}}{1+\delta_{\rm Kep}}
\end{equation}
and for the observational value $\delta_{\rm Kep}\sim 0.01$ we obtain $r_{\rm in}\sim 
r_{\rm out}\sim r_{\rm m}\sim 0.2$ pc, which yields an evaporation time given by
(\ref{evaporation}) consistent with the age of the galaxy NGC 4258.  The structure of the
supermassive cluster at the centre of the galaxy consists of a massive core and mantle with a
tenuous gas envelope. We can conclude that such a compact object should not produce a 
deviation of the Keplerian rotation curve which exceeds the observed value
$\Delta v\leq 3\, {\rm km}\,{\rm s}^{-1}$.

Let us now consider the physical collision time-scale. This can be estimated from the 
formula (Binney \& Tremaine 1987):
\begin{equation} 
t_{\rm coll}=\biggl[16\pi^{1/2}n\sigma r_{*}^2
\biggl(1+\frac{Gm}{2\sigma^2r_{*}}\biggr)\biggr]^{-1},
\end{equation}
where $n$ is the number density of stars, $r_{*}$ is the radius of the star, $m$ is the mass of the
star, and $\sigma$ is the velocity dispersion. For zero-temperature brown dwarfs and low-mass 
stars the mass-radius relation can be taken to be (Zapolsky \& Salpeter 1969; Stevenson 1991):
\begin{equation}
r_{*}=2.2\times 10^9\biggl(\frac{m}{M_{\odot}}\biggr)^{-1/3}
\biggl[1+\biggl(\frac{m}{0.0032M_{\odot}}\biggr)^{-1/2}\biggr]^{-4/3}\quad {\rm cm}.
\end{equation}
From (\ref{totalmass}) and (\ref{radius}), we can estimate the core density to be
\begin{equation}
\rho_0=\frac{12.93M}{R^3}.
\end{equation}
For $R\sim 0.2$ pc, $M\sim 3.6\times 10^7M_{\odot},m\sim 1.4M_{\odot}, r_{*}\sim
1.97\times
10^9$ cm and $\sigma\sim 1500\,{\rm km}\, s^{-1}$, we get $\rho_0\sim 5.9\times
10^{10}\,M_{\odot}\,{\rm pc}^{-3}$, $n=\rho_0/m\sim 4.2\times 10^{10}\,
{\rm pc}^{-3}$ and
\begin{equation}
t_{\rm coll}\sim 1\, {\rm Gyr},
\end{equation}
which is long enough to offset a rapid evolution of the massive stellar cluster through
coalescence of stars. Of course, as we decrease the radius $R$ of the massive core, then the 
time-scale of physical collisions of stars will decrease and lead to an unstable configuration.

In a recent article (Moffat 1997) the fate of a dense cluster of stars which is undergoing a
final stage of gravithermal catastrophe (Lynden-Bell \& Wood 1968) was analysed. Since this
phase of the evolution of a 
superdense core of stars is far from thermodynamic equilibrium, nonlinear cooperative
contributions
are expected to be important in the transport equations describing the last stage of evolution. It
was found that such nonlinear contributions can prevent the core redshift from increasing 
without limits as the core becomes increasingly dense, preventing the collapse to a black hole. 
In particular, the redshift can remain less than the critical value for relativistic collapse, resulting
in a stable, massive dark object at the center of a galaxy with a Newtonian core, mantle and thin
halo.

\section{Conclusions}

We see that it is not possible to rule out the hypothesis that NGC 4258 or other supermassive
galaxies are dense clusters of star-like objects with a mass $m\sim 1\,M_{\odot}$, since the
structure of the theoretical mass profile of such superdense clusters is not known with certainty.
Therefore, we must conclude that there is
presently no conclusive evidence for the existence of black holes at the center of galaxies such as
M31, M32, M87, NGC 4594, NGC 4258 and other potential black hole candidates. However, if
these dark massive objects with masses in excess of $10^6\,M_{\odot}$ are superdense
clusters of stars with relatively large central redshifts, then they would be of
considerable theoretical interest to the astrophysics community. 

Since it is difficult in the forseeable future to obtain an observational spatial resolution less than
$10^4$ Schwarzschild radii, it is not clear how the dark object at the center of galaxies
can be proved to be a black hole. Black holes may exist at the center of galaxies but it cannot be
claimed without further conclusive evidence that they have been been detected
by current observational data.

\acknowledgments

I thank the Natural Sciences and Engineering Research Council of Canada for the support of 
this work. 
\vskip 0.2 true in
\centerline{REFERENCES}
\vskip 0.3 true in
\noindent Binney, J., \& Tremaine, S. 1987, Galactic Dynamics  (Princeton: Princeton University 

\noindent Press)

\noindent Bisnovatyi-Kogan, G. S. \& Thorne, K. S. 1970, ApJ, 160, 875

\noindent Bisnovatyi-Kogan, G. S. \& Zel'dovich, Ya. B. 1969, Astofizika, 5, 223

\noindent Chandrasekhar, S. 1939, Stellar Structure (Dover Publications, New York).

\noindent Chandrasekhar, S. 1964, ApJ, 140, 417

\noindent Fackerell, E. D. 1970, ApJ, 160, 859

\noindent Fackerell, E. D., 1969, Ipser. J. R. \& Thorne, K. S. Comments Astrophys. and Space 

\noindent Phys., 1, 140 

\noindent Hoyle, F. \& Fowler, W. A. 1967,  Nature, 213, 373 

\noindent Ipser, J. R. 1969, ApJ, 158, 17

\noindent ------. 1970, Ap. And Space Sci. 7, 361

\noindent Ipser, J. R., \& Thorne, K. S. 1968, ApJ, 154, 251

\noindent Kormendy, J. 1993, in The Nearest Active Galaxies, ed. J. Beckman, L. Colina \& H.

\noindent  Netzer (Madrid: Consejo Superior de Investigaciones Cientificas), p. 197

\noindent Kormendy, J., \& Richstone, D. 1995, ARA\&A, 33, 581

\noindent Kormendy, J., et al. 1997, ApJ, 482, L139.

\noindent Lynden-Bell, D. \& Wood, R. 1968, R. MNRAS, 138, 495

\noindent Moffat, J. W. 1997, University of Toronto preprint astro-ph/9705258

\noindent Maoz, E. 1995, ApJ, 447, L91

\noindent Miyoshi, M., et al. 1995, Nature, 373, 127

\noindent Oppenheimer, J. R., \& Volkoff, G. M. 1939, Phys. Rev., 55, 374

\noindent Spitzer, L., \& Thuan, T. X. 1972, ApJ, 175, 31

\noindent Spitzer, L., \& Hart, M. H. 1971, ApJ, 166, 483

\noindent Stevenson, D. J. 1991, ARA\&A, 29, 163

\noindent Weinberg, S. 1972, Gravitation and Cosmology: Principles and Applications of the

\noindent  General Theory of Relativity (John Wiley \& Sons, Inc. New York), p. 325

\noindent Zapolsky, H. S. 1968, ApJ, 153, L163

\noindent Zapolsky, H. S., \& Salpeter, E. E. 1969, ApJ, 158, 809

\noindent Zel'dovitch, Ya. B., \& Poduretz, M. A. 1965, Astr. Zh., 42, 963 (English transl., 1966,

\noindent  Soviet Astr. - AJ, 9, 742)

\end{document}